\documentclass[12pt, a4paper]{iopart}
\usepackage[utf8]{inputenc}
\usepackage[square, comma, sort&compress, numbers]{natbib}
\usepackage{graphicx}
\usepackage{bm}
\usepackage[pdfpagemode={UseOutlines},
            bookmarks=true,bookmarksopen=true,
            bookmarksopenlevel=0,
            bookmarksnumbered=true,
            colorlinks,
            linkcolor={blue},
            citecolor={blue},
            urlcolor={blue},
            pdfstartview={FitV},
            unicode,
            breaklinks=true,
            plainpages=false
]{hyperref}

\expandafter\let\csname equation*\endcsname\relax
\expandafter\let\csname endequation*\endcsname\relax
\usepackage{amsmath}
\usepackage[version=3]{mhchem}

\usepackage{xspace}
\usepackage{textcomp}
\usepackage{textcomp}
\def\B{V$^2$/m$^2$Hz\xspace}

\usepackage{fancyhdr}
\renewcommand{\footnoterule}{%
  \kern -3pt
  \hrule width \textwidth height 0.4pt
  \kern 2pt
}

\begin{document}

\title{Electric-Field Noise above a Thin Dielectric Layer on Metal Electrodes}
\author{Muir Kumph$^1$, Carsten Henkel$^2$, Peter Rabl$^3$, Michael Brownnutt$^{4}$, Rainer Blatt$^{1,5}$}
\date{Sept 2015}
\address{$^1$ Institut f\"{u}r Quantenoptik und Quanteninformation
der \"{O}sterreichischen Akademie der Wissenschaften,
Technikerstrasse 21a, 
A-6020 Innsbruck, Austria}

\address{$^2$ Institute of Physics and Astronomy,
University of Potsdam,
Karl-Liebknecht-Str. 24/25,
D-14476 Potsdam, Germany}

\address{$^3$ Institute of Atomic and Subatomic Physics,
TU Wien,
Stadionallee 2,
A-1020 Vienna, Austria}

\address{$^4$ The University of Hong Kong, 
Pokfulam, 
Hong Kong}

\address{$^5$ Institut f\"{u}r Experimentalphysik,
Universit\"{a}t Innsbruck,
Technikerstrasse 25,
A-6020 Innsbruck, Austria}

\ead{muir.kumph@uibk.ac.at}

\begin{abstract}
The electric-field noise above a layered structure composed of a planar metal electrode covered by a thin dielectric is evaluated and it is found that the dielectric film considerably increases the noise level, in proportion to its thickness. Importantly, even a thin (mono) layer of a low-loss dielectric can enhance the noise level by several orders of magnitude compared to the noise above a bare metal.  Close to this layered surface, the power spectral density of the electric field varies with the inverse fourth power of the distance to the surface, rather than with the inverse square, as it would above a bare metal surface. Furthermore, compared to a clean metal, where the noise spectrum does not vary with frequency (in the radio-wave and microwave bands), the dielectric layer can generate electric-field noise which scales in inverse proportion to the frequency.  For various realistic scenarios, the noise levels predicted from this model are comparable to those observed in trapped-ion experiments. Thus, these findings are of particular importance for the understanding and mitigation of unwanted heating and decoherence in miniaturized ion traps.
\end{abstract}

\maketitle

\section{Introduction}
Electric-field fluctuations above metal surfaces are a common problem in many areas of physics and a severe limitation to precision measurements as diverse as space-based gravitational-wave detectors~\cite{Pollack:2008}, nano-cantilevers probing dispersion forces~\cite{Stipe:2001}, and the shielding of particle beams~\cite{Camp:1991}. In trapped-ion systems, electric-field noise at around 1~MHz and at distances of a few tens or hundreds of \textmu m from metallic electrodes significantly heats the ions~\cite{Brownnutt:2014}. This sets a limit on the coherence times that can be achieved in miniaturized trap designs which are currently developed for scalable quantum information processing. Ever since the observation of unexpectedly high heating rates~\cite{Monroe:1995_12} which could not be explained by the noise of the trapping circuitry, the role of the electric noise from surfaces in ion traps has attracted much experimental and theoretical attention. While a perfect conductor would not generate electric noise beyond the very low level of blackbody radiation, larger fluctuating electric fields are in principle expected above real conductors made of metals with non-vanishing resistive losses. However, early investigations~\cite{Lamoreaux:1997, Henkel:1999_06, Henkel:1999_11} showed that the noise levels expected from the metal's resistance are generally still far too low to account for the experimentally observed heating rates in ion traps. There are experimental indications that in some instances the high heating rates observed are related to conditions on the electrodes' surfaces. Various mechanisms have been proposed, including models based on fluctuating patch-potentials~\cite{Turchette:2000a, Dubessy:2009}, adatom dipoles, two-level fluctuators~\cite{SafaviNaini:2011}, or diffusing adatoms and charges~\cite{Stipe:2001, Henkel:2008}. Finding exactly which of these effects is significant in any given experiment, and whether other effects also play a role, constitutes an active area of experimental and theoretical research~\cite{Brownnutt:2014}. 

In this work the electric-field noise generated by a thin layer of a dielectric on top of a flat metal electrode is investigated. This scenario mimics surface conditions that are typically encountered in trapped-ion experiments: the surface of the metal electrodes, having been exposed to air and humidity, will usually be covered by a non-metallic layer such as native oxides or hydrocarbon compounds.  Recent experiments with trapped ions have indeed observed a considerable reduction of the electric-field noise after in-situ cleaning of the electrode surface with lasers~\cite{Allcock:2011}, ion-beam milling~\cite{Hite:2012} or plasma cleaning~\cite{McConnell:2015}. In this paper the contamination layer is modeled as a thin film with dielectric losses. By this means, analytic results for the spectral power of electric-field fluctuations $S_{E}$ at a distance, $d$, above the surface are calculated.
 
The analysis presented here shows that the presence of even a very thin dielectric (mono-layer) can increase the absolute level of electric-field noise by several orders of magnitude compared to a bare metal surface. It also shows that, for moderate distances from the surface $d > \delta$, where $\delta$ is the metal's skin depth, the distance dependence of the noise spectrum changes from a $d^{-2}$ to a $d^{-4}$ scaling.  Such a behavior is often attributed to localized surface potentials of microscopic origin, but arises here from a purely macroscopic description~\cite{Turchette:2000a, Dubessy:2009, SafaviNaini:2011}.  For many dielectric materials covering the electrodes, the permittivity $\epsilon$ and loss tangent $\tan\theta$ can be considered constant over a range of frequencies $\omega$~\cite{Jonscher:1977}, so that the power spectrum of the electric-field fluctuations decreases as $1/\omega$ with increasing frequency. The dielectric thickness and electrical properties could be measured independently with microwave loss spectroscopy~\cite{Swihart:1961, Garwin:1971, Winters:1991} or surface scanning probes, providing a more detailed test of this model.

The rest of this paper is organized as follows. The fluctuation--dissipation theorem is described in section~\ref{sec:fluctuations}, and a qualitative estimate is given using the method of image charges for the noise expected above a clean metal surface and above a metal covered with a dielectric. As a cross-check, in section~\ref{sec:thermalElectrodynamics} the noise spectrum is more rigorously computed for both a bare metal and a covered metal using methods of fluctuation electrodynamics.  The absolute levels of electric-field noise are given for common metals under realistic surface conditions in section~\ref{sec:commonMetals}.  Section~\ref{sec:discussion} discusses the relevance of the results and the outlook for experimentation in light of them.

\section{Electric-Field Fluctuations for a Charged Particle}
\label{sec:fluctuations}
\begin{figure}
\centering
\includegraphics[width=9cm]{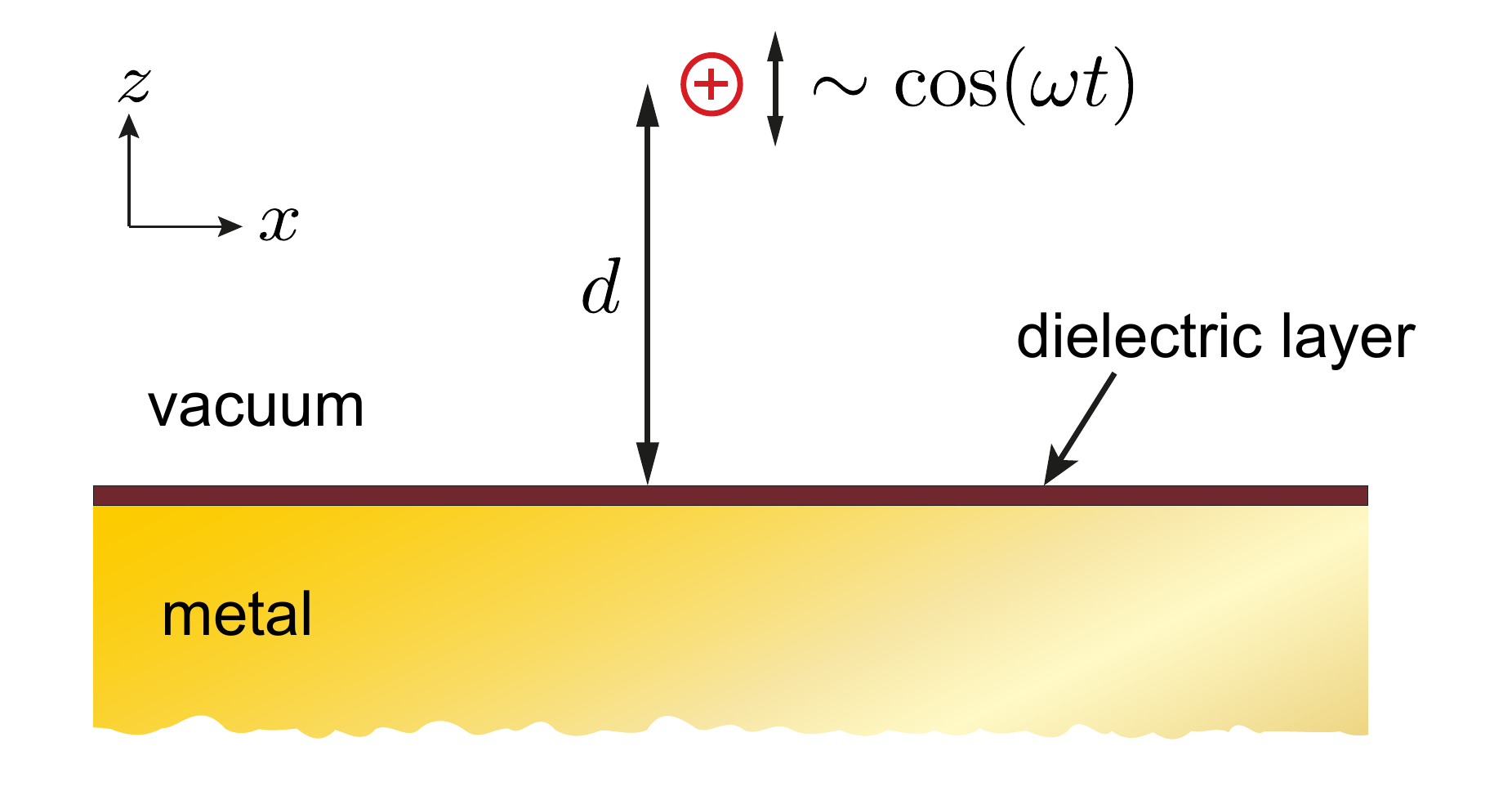}
\caption{A point-like particle is suspended in vacuum a distance $d$ above a conducting plane.  The plane is covered with material of thickness $t_{\rm d}\ll d$ characterized by a permittivity $\epsilon$ and loss tangent $\tan\theta$. The materials composing the structure are at temperature $T$.}
\label{fig:basicSetup}
\end{figure}

This paper considers a single charged particle (or ion) interacting with its surroundings as shown in figure~\ref{fig:basicSetup}.  The point-like ion is suspended in vacuum a distance, $d$, above a conducting plane.  At the surface of the plane, there is a material of thickness $t_{\rm d}\ll d$ characterized by a (real) permittivity $\epsilon$ and loss tangent $\tan\theta$. The materials composing the structure are at temperature $T$.

\subsection{Fluctuation--Dissipation Theorem}

Consider a single particle above a plane composed of some materials as in figure~\ref{fig:basicSetup}.  When the nearby materials have some non-zero temperature, they will transmit energy to the particle through fluctuating forces.  The motion of particle can also be damped by the surrounding materials via dissipative forces.  The fluctuation--dissipation theorem states that for a system composed of a single particle at equilibrium with its surroundings, at a temperature $T$, the energy that is transmitted to the particle by the surrounding material's fluctuating forces must be equal to the energy lost through dissipative forces to the environment~\cite{Einstein:1905Brownian, Nyquist:1928, Callen:1951, Lamoreaux:1997}.  The fluctuations from the surrounding materials are a property of the material's temperature $T$ and will affect the particle, even if the particle is no longer in equilibrium. Using the methods and notation outlined by Kubo~\cite{Kubo:2002}, the electric-field fluctuations above a metal surface, with and without a dielectric layer, are computed as follows. 

Consider a point-like particle moving in one dimension, where the dissipative force $F_\mathrm{d}$ is proportional to the speed of the particle, $u$, so that
\begin{equation}
F_\mathrm{d}= m\gamma u,
\end{equation}
where $m\gamma$ is the damping coefficient. More generally, this kind of formula applies in Laplace-Fourier space, with frequency-dependent $\gamma[\omega]$. This damping rate $\gamma[\omega]$ can be found by giving the particle an oscillatory motion at frequency $\omega$ and calculating the dissipated power due to this motion.  In addition to friction, the particle is subject to a random force of thermally activated origin. Of interest for us is the power spectrum of the force fluctuations, $S_\mathrm{F}( \omega )$. The convention used here is that of a single-sided power spectral density (PSD) (units of ${\rm N^2/Hz}$) which is given by
\begin{equation}
S_\mathrm{F}(\omega)= 2 \int_{-\infty}^\infty {\rm d}\tau \langle \delta  F(\tau) \delta F(0)\rangle  {\rm e}^{-{\rm i}\omega \tau},
	\label{eq:def-power-spectrum}
\end{equation}
where $\delta F(\tau)$ is the time-dependent variation of the force, $F$, from its long-term mean value.  The fluctuation--dissipation theorem links the fluctuating force to the dissipative damping, such that
\begin{equation}
S_\mathrm{F}(\omega)=4 k_\mathrm{B}T m\operatorname{Re} \gamma[\omega],
\label{eq:flucDiss}
\end{equation}
where $k_\mathrm{B}$ is Boltzmann's constant, and $\operatorname{Re} \gamma[\omega]$ is the real part of the damping rate. If the particle has a charge $q$, then the PSD of the fluctuating force is related to the power spectral density, $S_{E}$, of the electric-field fluctuations at the location of the particle by
\begin{equation}
\label{eq:SFtoSE}
S_{E}=\frac{S_{F}}{q^2}.
\end{equation}
The problem of computing the fluctuating electric field can thus be cast as a problem of calculating the dissipated power due to a forced motion of the charged particle.  In order to calculate the dissipation, the form of the electric field due to the charge above the surface is found.  The losses due to this electric field can then be computed. This is done for a clean metal surface in section~\ref{sec:metalLosses}, and for a system in which a thin dielectric covers the metal in section~\ref{sec:dielectricLosses}.

\subsection{Ohmic Losses in the Metal}
\label{sec:metalLosses}
\begin{figure}
\centering
\includegraphics[width=9cm]{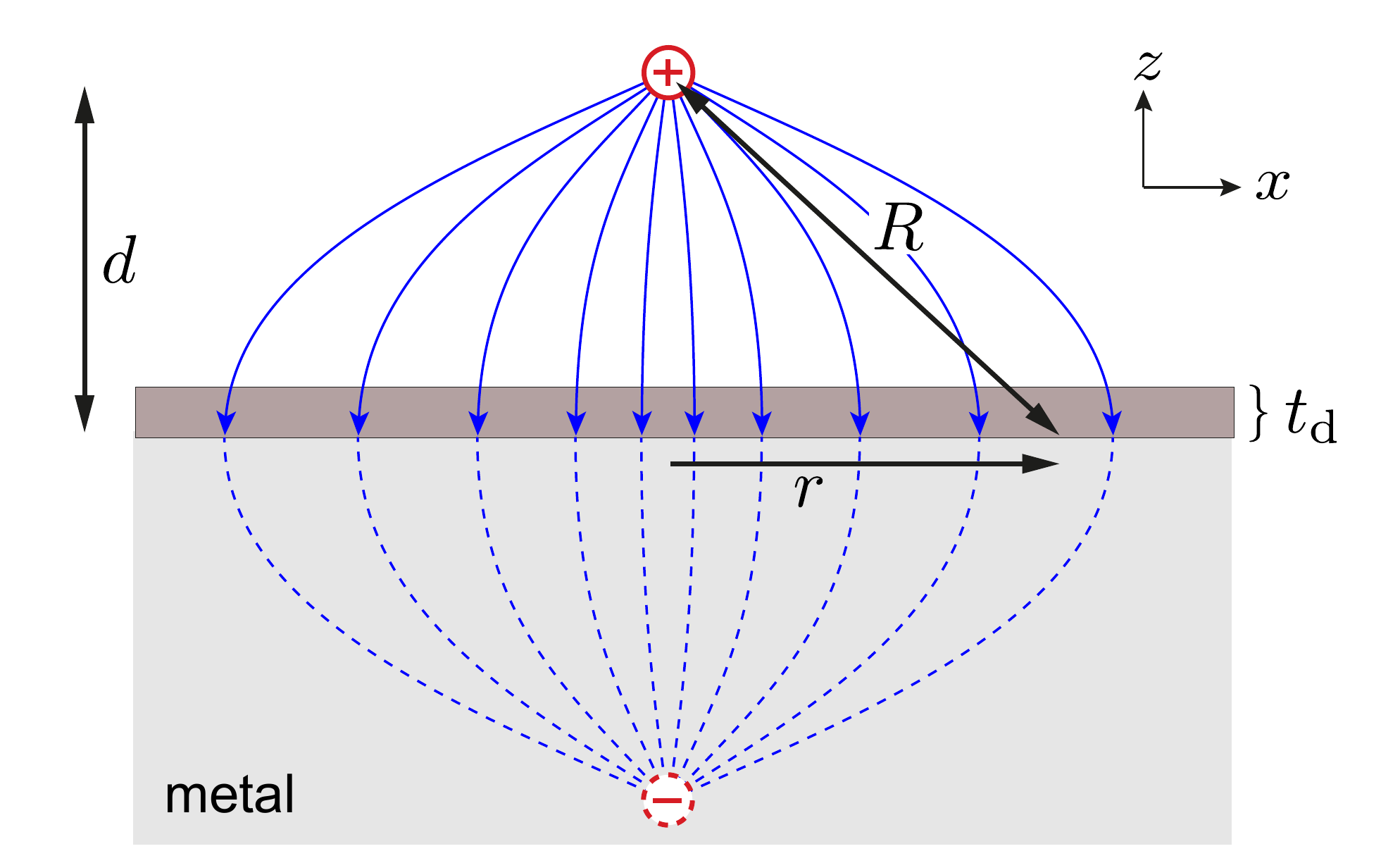}
\caption{For a single charged particle above the metal surface, the electric field above a metal surface forms a dipole pattern.  The method of electric images allows the $z$ component of the electric field at the metal surface due to the charged particle a distance $d$ above the surface to be easily calculated.}
\label{fig:chargeAboveMetal}
\end{figure}

The static electric field due to a charged particle above an ideal conductor is half of a dipole pattern. This is the same pattern as would arise in the upper half-space if two particles of charge $+q$ and $-q$ were separated by a distance $2d$, as shown in figure~\ref{fig:chargeAboveMetal}. This method of electric images~\cite{Kelvin:1872} allows the electric field at the conductor surface due to a charged particle a distance $d$ above the surface to be easily calculated. This is done by summing the fields of the real charge and its mirror charge.  In this case the electric field at the surface of the conductor only has a non-zero component of the electric field normal to the surface given by
\begin{equation}
E_z=-\frac{q d}{2\pi\epsilon_0 R^3},
\label{eq:surfaceField}
\end{equation}
where $\epsilon_0$ is the permittivity of free space and $R$ is the distance from the charged particle to the location on the surface.  The coordinates used here assume that the origin is located on the surface of the metal directly under the unperturbed charged particle, so that the $z$-axis goes through the particle.

The surface charge, $\sigma_\mathrm{s}$, present on an ideal conductor to produce the electric field at its surface is given by
\begin{equation}
\sigma_\mathrm{s}=\epsilon_0 E_z.
\end{equation}
If the charged particle is given a sinusoidal motion at frequency $\omega$, with a velocity-amplitude $\bm{u}$, the surface charge will be time-dependent and will produce a surface current, with amplitude $\bm{J}_\mathrm{s}$.  For motion normal to the surface of the conductor, only radial surface currents, with amplitude $J_{\mathrm{s}r}$, will be produced. Utilizing the continuity equation, these can be shown to be
\begin{equation}
J_{\mathrm{s}r}(r)=-\frac{1}{r}\int^r_0 {\rm d}r' r' \frac{\partial \sigma_\mathrm{s}}{\partial t} 
\,,
\end{equation}
where $r$ is the radial distance from the $z$ axis (i.e. $r^2=x^2+y^2$).  The term $\partial \sigma_\mathrm{s} / \partial t$ can be expressed as the time derivative of the electric field.  For a charged particle moving normal to the surface of the metal, with a small velocity amplitude, $|u_z| \ll d\omega$, this is
\begin{equation}
\frac{\partial \sigma_\mathrm{s}}{\partial t} = u_z \frac{q \left(2 d^2-r^2\right)}{2\pi R^5}.
\end{equation}

For an ideal conductor, the surface charges would respond instantly to the motion of the charge and reproduce the dipole pattern of figure~\ref{fig:chargeAboveMetal} at each moment in time.  However,  for materials with a non-zero resistivity, the induced surface currents produce an electric field parallel to the surface.  For metals commonly used to fabricate ion traps, such as copper, gold, and aluminum, and considering the case where the oscillating charge is about 100 \textmu m from the electrode, the resistance is so small that the field lines are not qualitatively different from figure~\ref{fig:chargeAboveMetal} up to frequencies in the THz band.  At higher frequencies the ideal conductor approximation breaks down and it becomes necessary to treat the metal more generally with a complex permittivity.  The analysis in this section is restricted to estimating the electric-field noise up to GHz frequencies with a distance between charged particle and surface greater than 100\,nm. This is the regime in which trapped-ion experiments operate, and is also relevant for many other experimental systems.

When the oscillating charge is much further away from the metal than the metal's skin depth, $\delta$, the current density in the conductor falls off exponentially with the distance from the surface~\cite{Jackson:1999SurfCurrents}.  In this sub-surface region, the amplitude of the radial current density, $j_r$, can then be approximated by a constant effective current density within the skin depth (i.e. $j_r=J_{\mathrm{s}r}/\delta$) and 0 elsewhere, so that
\begin{align}
j_r(z) & \approx \frac{q r u_z}{2 \pi  R^3 \delta } & & -\delta<z \le 0\\\notag
j_r(z) & \approx 0 & & z \le -\delta,
\end{align}
where $u_z$ is the amplitude of the $z$-component of the velocity of the particle. Within a metal of resistivity $\rho$, the cycle-averaged power-loss density $\langle p_\mathrm{loss} \rangle$, is then
\begin{equation}
\langle p_\mathrm{loss} \rangle=\frac{1}{2}\rho j^2_r,
\end{equation} 
where $\rho$ is the resistivity of the conductor.  By integrating the power-loss density over the volume of the whole conductor, this provides the total average dissipated power $\overline{P}_\mathrm{loss}$ in the conductor as a function of the amplitude of the $z$-component of the velocity $u_z$:
\begin{equation}
\overline{P}_\mathrm{loss}\approx\frac{q^2 \rho u^2_z}{16 \pi \, d^2 \delta }= \frac{1}{2}m u^2_z \operatorname{Re} \gamma[\omega] .
	\label{eq:power-loss-and-Re-gamma}
\end{equation}
From this, the real part of the damping rate can be obtained.  Using the fluctuation--dissipation theorem (see equation~(\ref{eq:flucDiss})), the electric-field spectrum is found. Far from the surface ($d>\delta$) this is
\begin{equation}
\label{eq:metalNoiseFar}
S^\mathrm{F}_{E,\perp}\approx\frac{k_\mathrm{B}T\rho}{2\pi \, d^2 \delta}.
\end{equation}

For currents flowing within a thin film of metal for which the thickness is less than the skin depth ($t_\mathrm{m} < \delta$), the current is confined to a smaller region than it would be in a bulk metal. This increases the losses, and the resulting electric-field fluctuations above such thin films are
\begin{equation}
\label{eq:metalNoiseThinFilm}
S^\mathrm{TF}_{E,\perp}\approx\frac{k_\mathrm{B}T\rho}{2\pi \, d^2 t_\mathrm{m}}.
\end{equation}

If the ion-electrode distance is smaller than both the skin depth and the metal's thickness ($d < \delta, t_\mathrm{m}$), then the currents (and electric fields) are confined even closer to the surface: to within a depth $\simeq d$~\cite{Wylie:1984}. The electric-field noise is then approximately
\begin{equation}
\label{eq:metalNoiseNear}
S^\mathrm{N}_{E,\perp}\approx\frac{k_\mathrm{B}T\rho}{2\pi \, d^3}.
\end{equation}

These results, obtained here by applying the fluctuation--dissipation theorem to a charge--image charge pair, reproduce essentially the same results derived independently by applying the fluctuation--dissipation theorem to a Green's function formalism of electrodynamics \cite{Agarwal:1975a, Henkel:1999_11}, which is further discussed in section~\ref{sec:bare-metal}.

The electric-field fluctuations above a metal due to resistive losses in the metal share many characteristics with fluctuations due to Johnson-Nyquist voltage noise of the electrodes and the connected circuitry.  The power spectrum is proportional to the resistivity of the electrical components and for ion-electrode separations greater than the skin depth, the power spectrum scales as $1/d^2$~\cite{Brownnutt:2014}.  However, one difference from voltage noise is that as the ion approaches the electrode to distances $d$ less than the skin depth $\delta$ (provided the electrode thickness, $t_{\rm m}$ is greater than $d$), the power spectrum scales as $1/d^3$.

\subsection{Losses in a Thin Dielectric Layer}
\label{sec:dielectricLosses}
In this section, the electric-field noise above a metal electrode covered with a thin layer of an isotropic dielectric with a thickness $t_{\rm d}\ll d$ is estimated.  The dielectric is characterized by a complex permittivity, $\varepsilon=\epsilon(1 + \rm{i}\tan\theta)$, a real permittivity $\epsilon$, and loss tangent $\tan\theta$.  It is further assumed here that the dielectric's loss tangent is not large ($\tan\theta < 1$) so that the electric-field pattern above the surface is still well approximated by a dipole pattern (see figure~\ref{fig:chargeAboveMetal}). 

The static energy density $w_0$ in the thin dielectric layer can be written as a function of the (real) static electric field $\bm{E}_0$ as,
\begin{equation}
w_0=\frac{1}{2}\epsilon \bm{E}_0 \cdot \bm{E}_0 = \frac{1}{2}\bm{D}_0\cdot \bm{E}_0,
\end{equation}
where $\bm{E}_0$ is the electric field in the dielectric layer due to the charged particle, 
and $\bm{D}_0=\epsilon\bm{E}_0$ is the (real) static electric displacement.

If there is a time-dependent change in the electric field, $\bm{E}(t)$, so that the total electric field is then $\bm{E}_\mathrm{total}(t)=\bm{E}_0+\bm{E}(t)$, the time dependent energy density, $w(t)$, is given by
\begin{equation}
\label{eq:ModEDensity}
w(t)=\frac{1}{2}\bm{D}_0\cdot \bm{E}_0 + \bm{D}_0 \cdot \bm{E}(t) + \frac{1}{2}\bm{D}(t) \cdot \bm{E}(t),
\end{equation}
where $\bm{D}(t)$ is the change in the electric displacement from its static value $\bm{D}_0$.  

Consider that the charged particle at a distance, $d$, above the surface undergoes a small-amplitude motion, $\delta\bm{r}(t)=\delta\bm{r}\cos(\omega t)$, at frequency $\omega$ and with amplitude $\delta\bm{r}$ ($|\delta\bm{r}|\ll d$), which produces a change in the electric field, $\bm{E}(t)$.  The first two terms on the right hand side of equation~(\ref{eq:ModEDensity}) 
therefore cycle-average to a constant or zero. Consequently, only the third term will 
contribute to the energy lost during a cycle of motion. 

The cycle-averaged rate of change of the energy density with time $\langle \partial w/\partial t \rangle$ is the time-averaged power loss density $\langle p_\mathrm{loss} \rangle$ in the dielectric,
\begin{equation}
\langle p_\mathrm{loss} \rangle = \langle \bm{E}\cdot \frac{\partial \bm{D}}{\partial t} \rangle.
\end{equation}
Using the complex formalism for the electric field, the power-loss density can be written as
\begin{equation}
\langle p_\mathrm{loss} \rangle = \frac{1}{2}\operatorname{Re}\left [\bm{\hat{E}}^*\cdot \frac{\partial}{\partial t}\bm{\hat{D}}\right ].
\end{equation}
where $\bm{\hat{E}}$ and $\bm{\hat{D}}$ are the complex amplitudes of the electric and displacement fields.  The complex amplitudes are defined by their relation to the time varying fields as,
\begin{align}
\bm{E}(t) &= \operatorname{Re}\left[ \bm{\hat{E}}\,{\rm e}^{-{\rm i}\omega t} \right]\\ \notag
\bm{D}(t) &= \operatorname{Re}\left[ \bm{\hat{D}}\,{\rm e}^{-{\rm i}\omega t} \right],
\end{align}
where $\omega$ is the frequency of the oscillations in the electric field and the complex amplitude of the displacement field is $\bm{\hat{D}=\varepsilon \bm{\hat{E}}}$. The cycle-averaged power loss density is then,
\begin{equation}
\label{eq:powerLossCycle}
\langle p_\mathrm{loss} \rangle=\frac{1}{2}\operatorname{Re}\left[-{\rm i}\omega \bm{\hat{E}}^*\cdot\bm{\hat{D}} \right]=\frac{\omega}{2}\epsilon\tan\theta|\bm{\hat{E}}|^2.
\end{equation}

If the motion of the particle is parallel to the surface of the metal in the $x$-direction with a small amplitude ($\delta x \ll d$), then the complex amplitude of the $z$-component of the electric field $\hat{E}_z$ at the surface can be expanded in $\delta x$ using equation~(\ref{eq:surfaceField}) as
\begin{equation}
\hat{E}_z= \frac{-3 q d}{2 \pi \varepsilon R^3}\frac{x \delta x}{R^2},
    \label{eq:Ez-oscillating-charge}
\end{equation}
where $x$ is the co-ordinate of the location on the layered surface below the charged particle and the factor $1/\varepsilon$ describes the dielectric screening in the material.  This approximation is equivalent to considering the oscillating charge as a dipole in the low-frequency limit.

The power density can then be computed as a function of $\delta x$. By integrating over the volume of the thin dielectric and averaging over a cycle (see eq.~\ref{eq:powerLossCycle}), the cycle-averaged power lost in the dielectric is found. The average power dissipated in the dielectric $\overline{P}_\mathrm{d}$ as a function of the amplitude of the oscillatory motion $\delta x$ is
\begin{equation}
\overline{P}_\mathrm{d}=\frac{3}{64\pi}
\frac{ \tan\theta }{ \epsilon (1 + \tan^2\theta) }
\frac{q^2 t_{\rm d} \omega (\delta x)^2 }{ d^4 } ,
\end{equation}
Using again the second equality in equation~(\ref{eq:power-loss-and-Re-gamma}) and knowing the amplitude $u_x = \omega \delta x$ of the particle velocity, the damping rate $m \operatorname{Re} \gamma[\omega]$ is found. Using equations~(\ref{eq:flucDiss}, \ref{eq:SFtoSE}), the spectrum of electric-field fluctuations parallel to the surface above the dielectric layer is
\begin{equation}
S^\mathrm{D}_{E,\parallel} = \frac{3}{8\pi}
\frac{ \tan\theta }{ \epsilon (1 + \tan^2\theta) }
\frac{k_\mathrm{B}T t_{\rm d}}{ \omega d^4 }.
	\label{eq:result-noise-parallel-1}
\end{equation}
This analysis can also be done for the dissipation of motion and electric-field fluctuations normal to the surface, for which the power spectrum due to the dielectric covering is
\begin{equation}
S^\mathrm{D}_{E,\perp} = \frac{3}{4 \pi }
\frac{ \tan\theta }{ \epsilon (1 + \tan^2\theta) } \frac{k_\mathrm{B}T t_{\rm d}}{\omega d^4}.
	\label{eq:result-noise-perpendicular-1}
\end{equation}

The noise due to the dielectric thin film occurs in addition to any noise due to the finite resistance of the metal plate itself (see equations~\ref{eq:metalNoiseFar}-\ref{eq:metalNoiseNear}).  However, as shown in section~\ref{sec:commonMetals}, for typical experimental parameter regimes, the noise from even very thin dielectric coatings (mono-layers) exceeds the noise due to resistive losses of the metal by several orders of magnitude, and so the noise from the metal can generally be neglected.

\section{Calculation from Fluctuation Electrodynamics}
\label{sec:thermalElectrodynamics}
In this section, the spectrum of the electric-field noise is calculated with the help of fluctuation electrodynamics in thermal equilibrium, using the fluctuation--dissipation theorem~\cite{Callen:1951,Agarwal:1975a}:%
\begin{equation}
S_{E,ij}( \bm{r}, \omega ) = 
\frac{ 4 k_\mathrm{B} T }{ \omega }
\mathop{\rm Im} G_{ij}( \bm{r}, \bm{r}; \omega ).
\label{eq:FDT-for-E}
\end{equation}
This classical approximation is valid because of the low-frequencies under consideration, $\hbar\omega \ll k_{\rm B} T$.
The spectrum, $S_{E,ij}$, gives the spectral expansion of the cross-correlation function, 
$\langle E_{i} E_{j} \rangle$, see equation~(\ref{eq:def-power-spectrum}). 
The Green tensor, $G_{ij}( \bm{r}, \bm{r}'; \omega )$, allows the electric field at the position of the trap centre, $\bm{r}$, radiated by a point dipole with complex amplitude $\hat{\bm{d}}$, located at $\bm{r}'$ and oscillating at a frequency $\omega$ to be calculated:
\begin{equation}
E_i( \bm{r}, t ) = 
\mathop{\rm Re} \Big[
\sum_{j}
G_{ij}( \bm{r}, \bm{r}'; \omega )
\hat d_j\, {\rm e}^{ - {\rm i} \omega t }
\Big].
\label{eq:def-E-Green-App}
\end{equation}
When evaluated at the metal plate [$\bm{r} = (x, y, z = 0)$] this recovers the field given in equation~(\ref{eq:Ez-oscillating-charge}). This field can be split into a free-space contribution and the reflection from the surface. Evaluating this in a frequency range where the distance to the surface is much shorter than the wavelength of the electric field, retardation can be neglected and the situation can be evaluated using electrostatics. The imaginary part of the reflection corresponds to the losses in the metal, relevant in equation~(\ref{eq:FDT-for-E}), and this yields, for fields parallel to the surface~\cite{Wylie:1984},
\begin{equation}
G_\mathrm{pp}( \bm{r}, \bm{r}; \omega ) \approx \frac{ 1 }{ 8\pi\epsilon_0 }
\int\limits_{0}^{\infty}\!{\rm d}k \,k^2\, R_\mathrm{r}( \omega, k )
\, {\rm e}^{ - 2 k d }.
\label{eq:parallel-Green}
\end{equation}
Here, $R_\mathrm{r}( \omega, k )$ is the surface's electrostatic reflection coefficient. If the latter is independent of $k$, then the integral with respect to $k$ in equation~(\ref{eq:parallel-Green}) can be simply performed%
\footnote{\[
\int\limits_{0}^{\infty}\!{\rm d}k \,k^2
\, {\rm e}^{ - 2 k d } = 
\frac{ 1 }{ 4 d^3 }
\]
}
and yields the field generated by an image dipole with amplitude $R_\mathrm{r}( \omega )$. 
According to the fluctuation--dissipation theorem expressed in equation~(\ref{eq:FDT-for-E}), the imaginary part of this image-dipole amplitude (related to dissipation in the surface) determines the electric-field noise (its fluctuation strength).

\subsection{Bare Metal}
\label{sec:bare-metal}

For a clean surface without contaminants, the reflection coefficient is $k$-independent~\cite{Kelvin:1872}, provided spatial dispersion (i.e. the anomalous skin effect) can be neglected. At distances greater than a few nanometers above the surface this is indeed the case and the method of image dipoles can be applied. 
\begin{equation}
\mbox{bare metal:} \qquad
R_\mathrm{r}( \omega, k ) = 
R_{\rm m}( \omega )
\,,\qquad
R_{\rm m}( \omega )
=
\frac{ \varepsilon_{\rm m}( \omega ) - 
\epsilon_0 }{ \varepsilon_{\rm m}( \omega ) + 
\epsilon_0 }.
\label{eq:bare-metal-R}
\end{equation}
This holds because, at low frequencies and for a good conductor, the complex dielectric function, $\varepsilon_{\rm m}$, is dominated by the conductivity $1/(\rho\omega)$, which is large compared to $\epsilon_0$. For example, the DC resistivity of gold typically exhibits $1/(\epsilon_0 \rho) \sim 10^{18}\,{\rm s}^{-1}$. Consequently,
\begin{equation}
\varepsilon_{\rm m}( \omega ) \approx 
\frac{ {\rm i} }{ \rho \omega } + \ldots .
\label{eq:eps-and-sigma}
\end{equation}
and to
a good approximation the dissipative part of the image dipole is
\begin{equation}
\mathop{\rm Im} R_\mathrm{r}( \omega, k ) \approx 
2 \mathop{\rm Im}\left( - \frac{ \epsilon_0 }{ \varepsilon_{\rm m}( \omega ) } \right)
\approx 2 \epsilon_0 \rho \omega
\ll 1
\label{eq:B}
\end{equation}
which is small, as expected for a good conductor. The field spectrum from equation~(\ref{eq:FDT-for-E}) becomes
\begin{equation}
\mbox{metal: } \qquad
S_{E,\parallel}( d, \omega ) \approx
\frac{ k_\mathrm{B} T \rho }{ 4 \pi\, d^3 }
\label{eq:result-metal}
\end{equation}
which is white. For the noise normal to the surface, a similar calculation~\cite{Wylie:1984} leads to a spectrum which is twice as large: $S_{E,\perp}( d, \omega ) = 2 S_{E,\parallel}( d, \omega )$, as also found in equation~(\ref{eq:metalNoiseNear}).

Note that the approximations used here do not reproduce a perfect conductor since they vanish in the limit $\rho \to 0$. For this case, retardation must be taken into account to capture the noise in the leading order. Explicit formulas can be found in Ref.~\cite{Henkel:1999_11}. It should also be noted that the short-distance approximation breaks down when $d$ becomes comparable to the skin depth in the metal: $d \sim \delta = [2\rho / (\mu \omega) ]^{1/2} \approx 75$\,\textmu m for gold at $1\,{\rm MHz}$ with a resistivity of 22.1 n\textohm$\cdot$m at a temperature of 293 K~\cite{Kaye:1995}, where gold's permeability $\mu=\mu_0$ is the vacuum permeability $\mu_0$. The $1/d^3$ scaling of equation~(\ref{eq:result-metal}) applies provided $d \ll \delta$. In the opposite limit, $d \gg \delta$, equation~(\ref{eq:result-metal}) must be multiplied by $2 d / \delta$, meaning that the noise exhibits a scaling of $\sim 1/d^2$~\cite{Henkel:1999_11}. This produces the same results obtained in section~\ref{sec:metalLosses}. For gold at $d = 100$\,\textmu m the noise level expected from equation~(\ref{eq:result-metal}) is $S_E \approx 10^{-17}$\,\B, much smaller than what is observed experimentally in ion traps~\cite{Brownnutt:2014}. Much larger noise levels can arise from covering layers as follows.

\subsection{Dielectric Covering Layer}
\label{sec:dielectricGreenFunc}
For a metal covered with a dielectric layer (thickness $t_{\rm d}$, 
complex permittivity $\varepsilon$), the (electrostatic) reflection coefficient is~\cite{Kelvin:1872,Yeh:book}
\begin{equation}
R_\mathrm{d}( \omega, k ) =
\frac{ R_{\varepsilon} + R_{\mathrm{m}\varepsilon}\, {\rm e}^{ - 2 k t_{\rm d} } }{
1 + R_{\varepsilon} R_{\mathrm{m}\varepsilon}\, {\rm e}^{ - 2 k t_{\rm d} } }
\,,
\label{eq:layer-R}
\end{equation}
where $R_{\varepsilon}$ and $R_{\mathrm{m}\varepsilon}$ are the reflection coefficients of the interfaces vacuum-dielectric and dielectric-metal respectively:
\begin{align}
R_{\varepsilon} &= \frac{ \varepsilon - \epsilon_0 }{ 
\varepsilon + \epsilon_0 }
\,, \\
R_{\mathrm{m}\varepsilon} &=
\frac{ \varepsilon_{\rm m}( \omega ) - \varepsilon }
{\varepsilon_{\rm m}( \omega ) + \varepsilon }.
\end{align}
The complex permittivity, $\varepsilon$, involves the loss tangent in its imaginary part, $\varepsilon = \epsilon ( 1 + \rm{i} \tan \theta)$.

Equation~(\ref{eq:layer-R}) can be approximated for the purposes of this analysis: from the integral in equation~(\ref{eq:parallel-Green}) it can be seen that the main $k$-vectors are $k = {\cal O}( 1 / d )$, so $k t_{\rm d} \ll 1$ for a thin layer. Combined with the assumption $|\varepsilon| =
{\cal O}( \epsilon_0 ) \ll |\varepsilon_{\rm m}|$, which is valid for a low-loss dielectric coating above a metal [see discussion above equation~(\ref{eq:eps-and-sigma})], 
a series expansion can be performed for the two small parameters $k t_{\rm d}$ and $\epsilon_0 / \varepsilon_{\rm m}$ to give
\begin{equation}
R_\mathrm{d}( \omega, k ) \approx 
1  
- 2 k t_{\rm d} \frac{ \epsilon_0 }{ \varepsilon } 
- 2 \frac{ \epsilon_0 }{ \varepsilon_{\rm m} }.
\label{eq:expanded-R-layer}
\end{equation}
Note the factor $k$ in the second term which, following integration with respect to $k$, leads to a different scaling with respect to distance, $d$.%
\footnote{\[
\int\limits_{0}^{\infty}\!{\rm d}k \,k^3
\, {\rm e}^{ - 2 k d } = 
\frac{ 3 }{ 8 d^4 }
\]
}
For a highly conductive substrate, this is also the dominating term in equation~(\ref{eq:expanded-R-layer}). The noise above a metal covered in a dielectric layer is thus equal to the sum of the noise from the dielectric layer and of the noise from the metal [given by equation~(\ref{eq:result-metal})]
\begin{equation}
\mbox{layer:}\qquad
S_{E,\parallel}^\mathrm{d}( d, \omega ) \approx
\frac{ 3 k_\mathrm{B} T t_{\rm d} }{ 8 \pi \epsilon_0 \omega \, d^4 }
\mathop{\rm Im}\left( - \frac{ \epsilon_0 }{  \varepsilon } \right)
+ \frac{ k_\mathrm{B} T \rho }{ 4 \pi\, d^3 }.
\label{eq:spectrum-layer}
\end{equation}
This can be rewritten in terms of the loss tangent and the DC permittivity, given that
\begin{equation}
\mathop{\rm Im}\left( - \frac{ \epsilon_0 }{ \varepsilon } \right)
=\frac{ \epsilon_0 \tan\theta }{ \epsilon (1 + \tan^2\theta) }.
\end{equation}
This method therefore independently reproduces the result of equation~(\ref{eq:result-noise-parallel-1}) which was derived by the methods of image charges. For some technical details and the extension of this calculation beyond electrostatics, see
the Appendix.

\section{Results For Common Electrode Materials}
\label{sec:commonMetals}
The model presented in section~\ref{sec:dielectricLosses} is quite general. It can be used to consider electrodes for which the dielectric covering is an intrinsic dielectric layer, such as a native oxide, as well as ones which are contaminated by some other non-conductive material. A thin dielectric layer covering ion trap electrodes has been measured on electrodes, which have significant electric-field noise with a level of approximately $10^{-11} \ldots 10^{-9}$\,\B at an ion-electrode separation $d\approx 50 \ldots 100$\,\textmu m~\cite{Hite:2012, Daniilidis:2014, Brownnutt:2014}.  The model presented here predicts comparable levels of noise for both contaminated gold electrodes and metals which form a native oxide such as copper.

Noble metals, such as gold, do not form oxides. Nonetheless, following exposure to air -- and particularly following the vacuum-bake process typically used in preparing trapped-ion systems -- the metal surface is typically covered with a few mono-layers of a dielectric substance such as hydrocarbons~~\cite{Hite:2012, Daniilidis:2014}.  The level of noise expected above a gold surface using the model presented here is estimated in section~\ref{sec:contaminationPSD}. Many metals develop a native oxide upon exposure to air and this native oxide can be a dielectric. This is the case for aluminium~\cite{Evertsson:2015}, niobium~\cite{Bach:2009PhD} and copper~\cite{Zheng:2002, Sarkar:2006}, all of which are standard materials for electrodes in ion traps. The level of noise expected above these metals (and their native oxides) is calculated in section~\ref{sec:nativePSD}. 

The levels of noise above metal electrodes with various dielectric coverings, calculated in sections~\ref{sec:contaminationPSD} and~\ref{sec:nativePSD}, can be compared to the level of noise above a bare metal. While the properties of the dielectric layers can vary significantly, the relevant properties of good metals are such that expected level of noise from the bare metal is relatively consistent between materials. In typical miniaturized surface ion-trap experiments the ion-electrode separation ($50\,$\textmu${\rm m} < d$) is about equal to the skin depth of the electrode material ($\delta \sim$ 50\,\textmu m) which is much greater than the range of electrode thicknesses used in miniaturized ion traps ($100\,\mathrm{nm}< t_{\rm m} <$ 10\,\textmu m). Using equation~\ref{eq:metalNoiseThinFilm} the electric-field noise expected 50\,\textmu m above a bare metal at 1\,MHz is approximately $10^{-16} \ldots 10^{-14}$\,\B. 

\subsection{Contamination}
\label{sec:contaminationPSD}
Metal electrodes can be contaminated with dielectric substances upon exposure to air.  For instance, a pure gold surface will be contaminated with at least 0.4\,nm of hydrocarbons (a mono-layer) within minutes of exposure to air~\cite{Smith:1980, Krim:1986, Vela:1990}. While the contamination on gold films exposed to air has been characterized to be largely hydrocarbon in nature with an approximate thickness of 0.4-2\,nm, the exact chemical structure and the radio-frequency electrical characteristics of these surface contaminants are not currently known.  

Consider, therefore, a gold electrode at room temperature with a 0.4\,nm thick hydrocarbon film on the surface having the electrical characteristics of a known hydrocarbon compound (pentane)~\cite{Kaye:1995}. This contamination would have a relative permittivity $\epsilon/\epsilon_0 \simeq 2$, with a loss tangent $\tan\theta \simeq 0.01$.   Using equation~\ref{eq:result-noise-parallel-1}, the power of the electric-field fluctuations 50\,\textmu m above the surface at 1\,MHz would be of order $10^{-11}$\,\B.

\subsection{Native Oxides}
\label{sec:nativePSD}
Many metals develop an oxide layer, called a native oxide, on any surface exposed to air and humidity. These oxides can form a dielectric a few nanometers thick and this is the case for metals commonly used in miniaturized ion traps, such as copper, aluminum and niobium.  The exact details of their thickness, chemical and and electrical properties can depend upon environmental conditions, as well as on the underlying metal. In some instances it is possible to reduce the electric-field noise experienced by trapped atomic ions above metallic electrodes by modifying the surface of electrodes which have a native-oxide layer~\cite{Daniilidis:2014, McConnell:2015}.  

In this section native oxides covering their associated metals are considered.  For each native oxide, the relative permittivity $\epsilon/\epsilon_0$, loss tangent $\tan\theta$ and thickness $t_\mathrm{d}$ is estimated.  And from these parameters, the corresponding power spectrum of electric field is provided using equations~\ref{eq:result-noise-parallel-1} and~\ref{eq:result-noise-perpendicular-1}. In each instance the noise at a distance of 50\,\textmu m above a planar surface at 300~K is calculated.

The alumina layer that forms as a native oxide on the surface of aluminum typically has a thickness, $t_{\rm d} \approx 4$~nm~\cite{Evertsson:2015}, a relative permittivity, $\epsilon/\epsilon_0 \simeq 8.5$~\cite{Kaye:1995}, and a loss tangent, $\tan\theta \simeq 0.001$~\cite{Auerkari:1996}. From equations~\ref{eq:result-noise-parallel-1} and~\ref{eq:result-noise-perpendicular-1}, the expected electric-field noise 50\,\textmu m above an aluminum surface with a native oxide at 1\,MHz is approximately $0.5\times10^{-12}$\,\B parallel to the surface and $1\times10^{-12}$\,\B normal to the surface.  

Niobium oxides have widely varying properties depending upon the exact stoichiometric ratio, crystal structure and test conditions~\cite{Bach:2009PhD}. For illustration, a 5~nm thick layer of \ce{Ni2O5} with a relative permittivity, $\epsilon/\epsilon_0\simeq41$, and a room-temperature loss tangent, $\tan\theta\simeq0.01$ is considered here~\cite{Emmenegger:1968, Brunner:1968}.  Again from equations~\ref{eq:result-noise-parallel-1} and~\ref{eq:result-noise-perpendicular-1} the expected power spectral density (PSD) 50\textmu m above the surface at 1\,MHz is around $1.5\times10^{-12}$\,\B parallel to the surface and $3\times10^{-12}$\,\B normal to the surface. 

Copper oxides also have widely varying properties depending on exactly how they are produced. They tend to have large relative permittivities and high losses~\cite{Zheng:2002, Sarkar:2006}. Their thickness grows over time on exposure to the humidity in air without limit. Because of these wide variations it is hard to give a general level of expected noise. However, for illustration, a 5~nm thick layer of \ce{CuO} with a relative permittivity, $\epsilon/\epsilon_0\simeq20$, and loss tangent, $\tan\theta\simeq0.5$, is considered here. From equation~(\ref{eq:result-noise-parallel-1}) the expected electric-field noise 50\textmu m above the surface at 1\,MHz is of order $10^{-10}$\,\B.  

\subsection{Distance and Frequency Scaling of Common Materials}
Our analysis presented in sections~\ref{sec:fluctuations} and~\ref{sec:thermalElectrodynamics} has shown that a thin dielectric layer can significantly modify the electric field noise spectrum and change its scaling with distance and frequency. For a situation where the ion-surface separation $d$ is much larger than the dielectric layer, $t_{\rm d}$ as well as the thickness of the current layer in the metal (either $t_{\rm m}$ or $\delta$), the distance scaling changes from $d^{-2}$ to $d^{-4}$. From the simple model described in section~\ref{sec:dielectricLosses}, this result follows from the fact that in a dielectric layer local losses scale as $|E_z(t)|^2$, which falls much faster, with increasing $d$, than the radial current density squared, $j_r^2(t)$, responsible for resistive losses in the metal (see sec.~\ref{sec:metalLosses}).  Compared to the bare metal, the result for $S_E(\omega)$ for a dielectric layer contains another factor of $\omega^{-1}$, but in general also the frequency and temperature dependence of the dielectric loss must be taken into account and
\begin{equation}
S_E(\omega) \sim  \frac{ T \, \mathop{\rm Im} \varepsilon(\omega; T)}{\omega}.
\end{equation}   
For a simple Debye model for the dielectric constant, $\varepsilon(\omega)\approx \epsilon/(1+i\omega \tau)$, where $\tau$ is a characteristic damping time, one would obtain $S_E(\omega)=const.$ for $\omega\ll \tau^{-1}$ and $S_E(\omega)\sim 1/\omega^2$ for $\omega\gg \tau^{-1}$. However, it is know that most real materials have a much weaker frequency dependence in the RF to microwave frequency regime~\cite{Jonscher:1977} and therefore, depending in detail on the dielectric material, a scaling $S_E(\omega)\sim \omega^{-\alpha}$, with $\alpha\sim 1$ is expected.  

The temperature dependence of the complex permittivity of materials varies widely.  However, for materials whose permittivity does not change substantially with temperature, the noise would scale linearly with temperature $T$.  For the native oxides of aluminum and niobium, the loss tangent tends to decrease with temperature~\cite{Auerkari:1996}.  In general, microwave and radio-frequency spectroscopy with conventional tools, or using a trapped ion as a probe could be used to infer the temperature dependence of the complex permittivity.

\begin{figure}
\centering
\includegraphics[width=8cm]{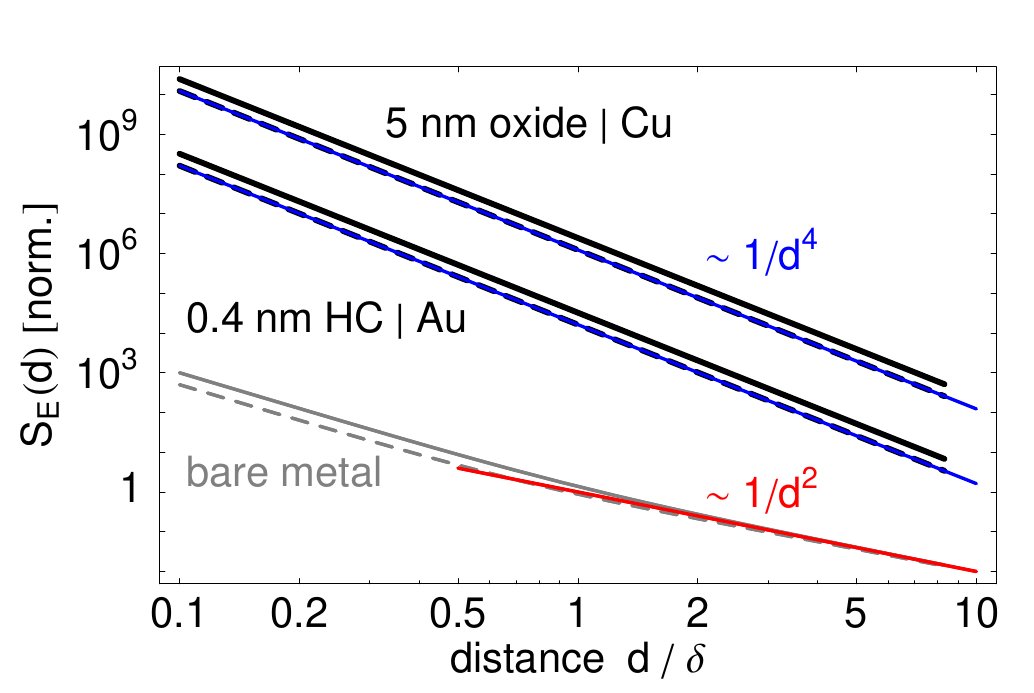}
\caption{The normalized electric-field-noise-levels and distance scaling for a bare gold or copper electrode, a gold electrode with 0.4\,nm hydrocarbon (H.C.) contamination and a 5\,nm film of copper oxide on a copper electrode.  Dashed lines: fields parallel to the surface; solid lines: fields normal to surface.  The noise above a bare metal scales as $1/d^2$ when the distance $d$ between the charged particle and the metal surface is larger than the skin depth $\delta$.  When $d$ is smaller than the skin depth, the scaling changes to $1/d^3$.  Even very thin layers of dielectric materials covering the metal electrodes will produce an electric field noise above the surface, which is orders of magnitude above that produced from a metal and scales as $1/d^4$. Reference noise level and skin depth are $S_E=1\times10^{-16}$\,\B, $\delta = 75$\,\textmu m.}
\label{fig:noiseComparison}
\end{figure}

Figure~\ref{fig:noiseComparison} shows the normalized electric-field-noise levels vs. distance for a bare gold or copper electrode, a gold electrode with 0.4\,nm hydrocarbon (HC) contamination and a 5\,nm film of copper oxide on a copper electrode. For thick metal electrodes ($t_\mathrm{m} > d$), the noise above a bare metal scales as $1/d^2$ [as $1/d^3$] when the distance $d$ between the charged particle and the metal surface is larger [smaller] than the skin depth $\delta$, respectively.  Even very thin layers of common dielectric materials covering the metal electrodes will produce an electric field noise above the surface, which is orders of magnitude above that produced from a metal and scales as $1/d^4$. Assuming the loss tangent $\tan\theta$ and permittivity $\epsilon$ are essentially constant~\cite{Jonscher:1977} the expected power spectrum is inversely proportional to the frequency $\omega$.

\section{Outlook and Summary}
\label{sec:discussion}
Numerous mechanisms have previously been put forward to account for the electric-field noise observed in miniaturized ion traps above conductors. The challenge is to match the observed levels of noise which are well above those predicted for bare metals, and their scaling with relevant parameters like distance, frequency, and temperature. Numerous experiments have been performed to characterise the noise, often with apparently-conflicting results. Taken together the results seem to point to the fact that different experiments are limited by different, possibly multiple, sources of noise \cite{Brownnutt:2014}. 

The mechanism considered in this paper is by no means a panacea to explain all experimental observations. Rather it is to be added to the list of noise sources which must be considered (and if necessary eliminated) in any given experiment. Thin dielectric coatings that cover a metallic electrode have been analyzed here and it is found that electric-field fluctuations many orders of magnitude stronger than above a clean metal surface are to be expected.  This is consistent with a number of experimental results, which reduced the electric-field fluctuations by modifying the surface.  For instance, it has been shown that treatments which altered the native oxide of superconducting cavities were able to improve the quality factors of such cavities~\cite{Garwin:1971}.  In ion traps, laser-ablation cleaning has been seen to cause a slight reduction in the electric-field noise above aluminum electrodes~\cite{Allcock:2011}, and plasma cleaning has been used to reduce the electric-field noise above niobium electrodes \cite{McConnell:2015} and copper/aluminum electrodes~\cite{Daniilidis:2014}.

For metals such as gold, which does not support a native oxide, the analysis presented here shows that even mono-layers of dielectrics which adhere to a non-passivated gold surface exposed to air will produce substantial electric-field noise. This is consistent with experimental results which show that argon-ion cleaning of gold electrodes can significantly reduce the electric-field noise above such surfaces~\cite{Hite:2012}.  The model presented here could be tested in detail with setups~\cite{Daniilidis:2014} where a controlled surface coating is deposited on the trap electrodes. One would expect a difference between islands and continuous films, amorphous or annealed.  Alternatively, the electric properties of surface layers may be tested with microwaves whose fields are confined to the sub-surface region by the skin effect. Similar techniques have been applied for superconducting cavities~\cite{Garwin:1971}. More generally, the crucial role of electrode coatings put forward here may help to understand why some traps develop increased anomalous heating over time (``aging"), while others perform well over periods of months.

Noise of the type modelled here can be distinguished from other noise sources. For instance, in trapped-ion systems, if Johnson-Nyquist noise is the dominant source of fluctuating electric fields, this will predominantly originate in the attendant electronics in the system, rather than the ion-trap electrodes themselves. Consequently the noise level varies as a function of the temperature of the electronics. In contrast, noise due to dielectric coverings on the electrodes varies as a function of the electrode temperature, which can be controlled independently of the attendant electronics.

In addition to highlighting a possible source of noise in trapped-ion experiments, the analysis of this paper suggests a novel method of reducing the electric-field noise in experiments. If there is an existing dielectric layer on the electrodes, it could be modified to increase its (real) permittivity $\epsilon$ or reduce its loss tangent $\tan\theta$.  This would reduce the electric-field noise (see equation~\ref{eq:result-noise-parallel-1}).  For example, copper electrodes exposed to air will invariably have a layer of copper oxide \ce{CuO} on them.  Copper oxide can transition, by means of a temperature treatment, to a giant permittivity material with a relative permittivity $\epsilon/\epsilon_0 \simeq 10^4$~\cite{Sarkar:2006}.  For metals such as gold, which are easily contaminated upon exposure to air, it may be possible to mitigate contamination through passivation of the bare metal by a thin film of a substance with a large permittivity during fabrication.  For instance, a film of a ceramic such as \ce{SrTiO3} with a relative permittivity $\epsilon/\epsilon_0 \simeq 10^4$ may provide a suitably high dielectric screening and passivation~\cite{Wang:2014}.

The simple model presented here of an infinite sheet of conductor with a uniform layer of a dielectric coating could be extended to include other situations.  For instance, the expected electric-field noise for three-dimesional electrodes or non-uniform patches of various materials could be calculated with the same basic theory.  It is expected that the distance scaling would depend upon the geometry of the electrodes~\cite{Brownnutt:2014} and patches of high-loss materials would increase the electric-field noise locally~\cite{Daniilidis:2011}.  Such customization of the theory presented here would allow for the model to be applied to more specific experimental situations.  

In summary, this paper describes how the expected thermal noise above metal electrodes coated with various dielectric materials can be calculated using a simple macroscopic model. It is shown that native oxides of common metals and mono-layers of hydrocarbon contamination can produce levels of electric-field noise which could be of concern to a number of experiments.

\section*{Acknowledgements}
The authors thank Nikos Daniilidis, Yves Colombe, Philipp Schindler, Ron Folman, Baruch Horovitz, and Ferdinand Schmidt-Kaler for discussions. This work was supported by the Austrian Science Fund (FWF) via the project Q-SAIL, the SFB FOQUS, the START Grant Y 591-N16 and the Institut f\"ur Quanteninformation GmbH. C.H. acknowledges support from  the DFG through the DIP program (FO 703/2-1).

\newpage
\appendix
\section*{Appendix: Efficient calculation of electric noise spectra}
\setcounter{section}{1}
\label{sec:appendixA}

\subsection{Introduction}

The experimental conditions typical for ion traps can
be summarized as
\begin{itemize}
\item low frequencies
$\hbar \omega \ll k_{\rm B} T$
\item good metallic conductors
$\varepsilon_m( \omega ) \approx {\rm i} / (\rho \omega)$
\item trap--surface distance small compared to wavelength
$d \ll c / \omega$
\item but comparable to skin depth
$\delta = [2 \rho / (\mu_0 \omega)]^{1/2} \sim d$
\item thin dielectric coating with thickness 
$t \ll d, \delta$
\end{itemize}
Under these conditions, the fluctuation--dissipation theorem
can be simplified because of the low frequencies involved. 
The spectral correlation function of the electric
field is used here for positive frequencies only. 
Attention: this convention for the noise spectrum is a 
factor $2$ larger than in other papers~\cite{Agarwal:1975a}
\begin{equation}
S_{E,ij}( {\bf r}, \omega ) = 
\frac{ 4 k_\mathrm{B} T }{ \omega }
\mathop{\rm Im} G_{ij}( {\bf r}, {\bf r}; \omega ).
\label{eq:FDT-for-E-App}
\end{equation}
The Green tensor is defined (and normalized) according
to electric dipole radiation by equation~(\ref{eq:def-E-Green-App}) 
\begin{equation}
E_i( \bm{r}, t ) = 
\mathop{\rm Re} \Big[
\sum_{j}
G_{ij}( \bm{r}, \bm{r}'; \omega )
\hat d_j\, {\rm e}^{ - {\rm i} \omega t }
\Big].
\label{eq:def-E-Green}
\end{equation}
for a point dipole with complex amplitude $\hat{\bf d}$, frequency $\omega$,
located at position ${\bf r}'$.

\subsection{Green tensor}

We work in a planar geometry: expansion in plane waves with wave vector
${\bf k}$ parallel to surface~\cite{Agarwal:1975a,Wylie:1984}.
For ${\bf r} = {\bf r}'$,
only the diagonal components of the Green tensor
are nonzero. It splits naturally in two contributions:
free space radiation and reflection from the surface.
The first part is the same as if the surface were at infinite distance; it gives an imaginary part of
\footnote{adapted from Eqs.(11, 12) in \cite{Henkel:1999_11}; attention:
the correlation function there is not symmetrized}:
\begin{equation}
\mathop{\rm Im} G_{ij}^{\infty}( {\bf r}, {\bf r}; \omega ) =
\delta_{ij} \frac{ \omega^3 }{ 6 \pi \varepsilon_0 c^3 }.
\end{equation}
The reflection from the surface gives the distance ($d$) dependent
parts for fields parallel ($p$) and normal ($n$) to the surface
[from Eq.(14) in \cite{Henkel:1999_11}]:
\begin{eqnarray}
\mathop{\rm Im} G_{pp}( z, \omega ) &=& 
\frac{ 1 }{ 8 \pi \varepsilon_0 }
\int\limits_{0}^\infty \!
{\rm d}k \, {\rm e}^{ - 2 k d }
\left( 
	\frac{ \omega^2 }{ c^2 } 
	\mathop{\rm Im} r_s( k ) 
+ 
	k^2 \mathop{\rm Im} r_p( k ) 
\right)
\nonumber
\\
\mathop{\rm Im} G_{nn}( z, \omega ) &=& 
\frac{ 1 }{ 4 \pi \varepsilon_0 }
\int\limits_{0}^\infty \!
{\rm d}k \, {\rm e}^{ - 2 k d }
k^2 
\mathop{\rm Im} 
r_p( k ).
\label{eq:expansion-G}
\end{eqnarray}
We have used here 
the non-retarded approximation, applying the general rules: relevant 
$k$-vectors much larger than $\omega/c$. This is because their
typical size is (from the exponential) $k \sim 1/d \gg \omega/c$.
This applies to fields above the
surface (outside the dielectric): they correspond
to electrostatic fields with $\nabla^2 {\bf E} = 0$, hence
the normal wavevector $k_z = ( \omega^2 / c^2 - k^2)^{1/2}$
is approximately purely imaginary, $k_z = {\rm i} k$.
We see that if the s-polarization does not contribute, the
normal field spectrum is a factor $2$ above that for parallel fields.

\subsection{Reflection coefficients}

These are given by the Fresnel formulas. For a simple interface $12$ 
(field incident from
medium $1$), they depend on normal wave vectors $q_a$ in the media
($a = 1,2$)
\begin{equation}
q_{a} = ( \varepsilon_a \mu_0 \omega^2 - k^2 )^{1/2},
\label{eq:def-medium-q}
\end{equation}
where $\varepsilon_a$ is the dielectric function ($\varepsilon_0$
for vacuum). The two principal polarizations (also denoted
TM, TE instead of $p$, $s$) yield:
\begin{eqnarray}
r_p^{12} &=& 
\frac{ \varepsilon_2 q_{1} - \varepsilon_1 q_{2} }%
{ \varepsilon_2 q_{2} + \varepsilon_1 q_{2} }
\\
r_s^{12} &=& 
\frac{ q_{1} - q_{2} }%
{ q_{1} + q_{2} }.
\end{eqnarray}
For a layered system (media $012$ from top to bottom, incidence
from `above')~\cite{Yeh:book}
\begin{equation}
r^{012} = 
\frac{ r^{01} + r^{12}\, {\rm e}^{ 2 {\rm i} q_1 t } }{
1 + r^{01} r^{12} \, {\rm e}^{ 2 {\rm i} q_1 t } },
\end{equation}
where $t$ is the thickness of the layer (medium $1$). This formula
has the same structure for both polarizations (the planar geometry
preserves the two principal polarizations).

Our goal is now: simplify these expressions
without compromising too much accuracy as
appropriate for our parameters. Assumption: dielectric
function of layer is comparable to vacuum, while 
below there is a good conductor. This gives
$\varepsilon_1 \sim \varepsilon_0 \ll \varepsilon_2$
(to be understood in absolute values, all medium dielectric 
functions are in general complex).

Using the non-retarded approximation, we expand the square roots
in Eq.(\ref{eq:def-medium-q}) and get the approximate medium wave 
vectors
\begin{eqnarray}
q_a \approx {\rm i} k 
	- \frac{ {\rm i} \varepsilon_a \mu_0 \omega^2 }{ 2 k }
\,, a = 0, 1\,, \qquad
q_2 \approx 
( 2 {\rm i} / \delta^2 - k^2 )^{1/2}
\,,
\end{eqnarray}
where the dielectric function of the conducting medium~2
has been re-written with the skin depth $\delta$:
\begin{equation}
\varepsilon_2( \omega ) \approx \frac{ {\rm i} }{ \omega \rho } 
= \frac{ 2 {\rm i} }{ \mu_0 \omega^2 \delta^2 }.
\end{equation}
Since we allow for $d \sim \delta$, we have to deal with
$k$-vectors $k \sim 1/\delta$ and we do not expand $q_2$. 

\subsubsection{s-Polarization}

This polarization is often ignored in the non-retarded limit.
We start with it because the discussion is somewhat simpler.
In the leading order for the vacuum--dielectric interface
\begin{equation}
r^{01}_s \approx 
\frac{ (\varepsilon_1 - \varepsilon_0) \mu_0
\omega^2 }{ 4 k^2 } 
=
\left( \frac{ \varepsilon_1  }{ \varepsilon_0 } - 1  \right)
\frac{ \omega^2 }{ 4 k^2 c^2 }
\ll 1
\end{equation}
while for the dielectric-metal interface, we have 
${\cal O}(1)$ reflection:
\begin{equation}
r^{a2}_s \approx
\frac{ {\rm i} k - q_2 }{ {\rm i} k + q_2 }
\,,\qquad
a = 0, 1
\,.
\end{equation}
This is not unitary since $q_2$ is in general complex.
For the layered geometry, we neglect the 01 reflection compared 
to 12 and get the following approximation
\begin{equation}
r_s^{012} \approx 
\frac{ {\rm i} k \delta - (2{\rm i} - k^2\delta^2)^{1/2} }{
{\rm i} k \delta + (2 {\rm i} - k^2\delta^2)^{1/2} }
 \, {\rm e}^{ - 2 k t },
\end{equation}
which is just a factor ${\rm e}^{ - 2 k t }$ smaller than
the reflection from the bare metal surface. (The exponential takes
into account that the `strong reflection' happens at the lower
interface, whose distance from the charge is $d + t$.) The 
dielectric losses in the thin layer do not appear in this
order. (This is OK as long as its dielectric function
$\varepsilon_1$ is much smaller than that of the metal.)

\begin{figure}[th]
\includegraphics*[height=45mm]{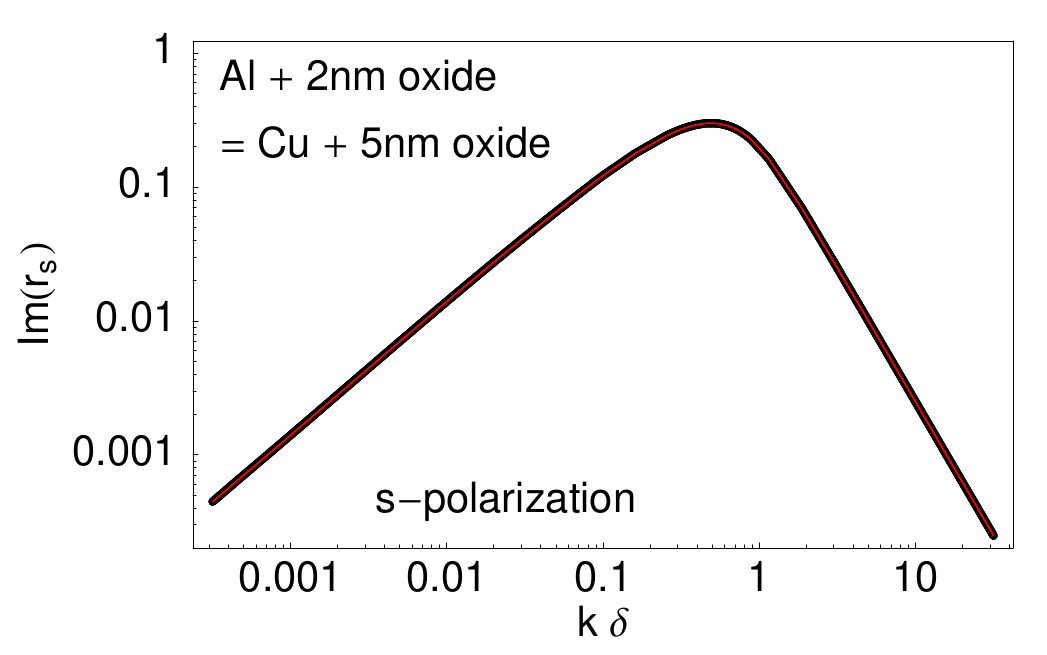}\hspace*{3mm}%
\includegraphics*[height=45mm]{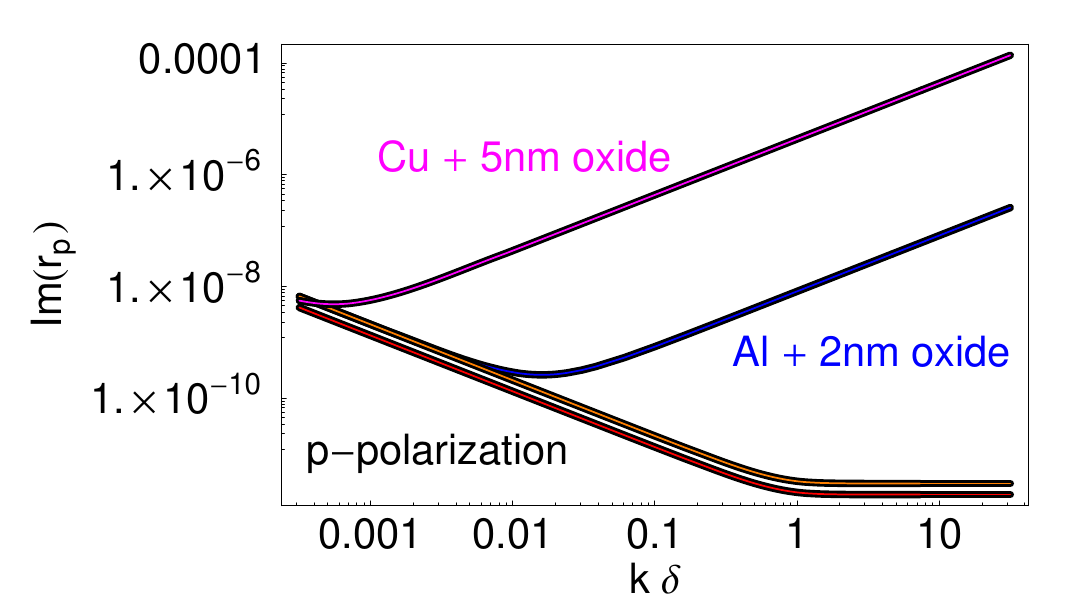}
\caption[]{%
Dielectric losses of covered metal, expressed
as imaginary part of reflection coefficients. The thick curves
give the exact calculation (including retardation), the thin 
(colored) ones the approximations discussed in the text.
\\
(\emph{left}) s-Polarization: bare and covered metals give
the same result. 
(\emph{right})
p-Polarization: upper lines with layer, lower lines bare metal.
Frequency
$\omega/2\pi = 1\,{\rm MHz}$.
Parameters for copper: conductivity from wikipedia
($\delta = 65\,{\rm \mu m}$), 
oxide layer with $t = 5\,{\rm nm}$, dielectric constant $20$,
loss tangent $0.5$.
For aluminium:
conductivity from wikipedia
($\delta = 75\,{\rm \mu m}$), 
oxide layer with $t = 2\,{\rm nm}$, dielectric constant $8.5$,
loss tangent $10^{-3}$. 
The skin depth $\delta$ is used to scale the $k$-vector.
With this scaling, the s-polarization gives a `universal
curve' that is independent of the material. Note that
in p-polarization, the losses scale with the thickness and
the loss tangent. 
}
\label{fig:Im-r-vs-k}
\end{figure}

Fig.\ref{fig:Im-r-vs-k}(left) shows that these approximations work 
very well for a broad range of $k$-vectors. As apparent from
Eqs.(\ref{eq:expansion-G}), the imaginary parts of the reflection
coefficients determine the noise spectrum; this is shown in the 
plots.

\subsubsection{p-Polarization}

The situation is somewhat reversed: at the vacuum-dielectric 
interface, we have ${\cal O}( 1 )$ reflection:
\begin{equation}
r^{01}_p \approx \frac{ \varepsilon_1 - \varepsilon_0 }
{ \varepsilon_1 + \varepsilon_0 },
\end{equation}
while at the dielectric-metal interface there is a (nearly) perfect 
reflector:
\begin{equation}
r^{a2}_p \approx 1 + 2 \frac{ (2{\rm i} - k^2\delta^2)^{1/2} }{ k \delta }
\varepsilon_a \omega \rho
\,,\qquad
a = 0,1
\,,
\end{equation}
whose imaginary part is small because
$\varepsilon_a \omega \rho = {\cal O}( 10^{-11} )$
for good conductors and MHz frequencies
(this is simply the ratio 
$\varepsilon_a / \varepsilon_2( \omega )$).
This small deviation
from perfect reflection is at the end responsible for the imaginary 
part.

Expansion of the layered reflection coefficient for
a thin layer $q_1 t \approx {\rm i} k t \ll 1$ gives
\begin{equation}
r^{012}_p \approx 
1 + 2 \frac{ (2{\rm i} - k^2\delta^2)^{1/2} }{ k \delta }
\varepsilon_0 \omega \rho
- 2 \frac{ \varepsilon_0 }{ \varepsilon_1 } k t.
\label{eq:rp-approx}
\end{equation}
Here, we have also expanded the exponential 
${\rm e}^{ 2 {\rm i} q_1 t }$.
Fig.\ref{fig:Im-r-vs-k}(right) shows that these approximations work 
very well for a broad range of $k$-vectors.

\subsection{Spectra}
Use scale factor $1/\delta$ to introduce a dimensionless
$k$-vector (see the products $k\delta$ in the formulas above).
Then by combining the FDT and the expansion of the Green function,
we find
\begin{eqnarray}
S_{E,p}(d, \omega ) &=& 
\frac{ k_{B} T }{ 2\pi \varepsilon_0 \omega \, \delta^3 }
\left[
s_p( d / \delta; \omega ) 
+
\frac{ \omega^2 \delta^2 }{ c^2 }
s_s( d / \delta ) 
\right]
\\
S_{E,n}( d, \omega ) &=& 
\frac{ k_{B} T }{ \pi \varepsilon_0 \omega \, \delta^3 }
s_p( d / \delta; \omega ),
\end{eqnarray}
where the dimensionless functions $s_s$ and $s_p$ are given
in Eqs.(\ref{eq:s-integral-approx}, \ref{eq:p-integral-approx-m},
\ref{eq:p-integral-approx-d}) below.

Observe the factor 2 between parallel and normal noise spectra,
apart from the contribution of the s-polarization. That one
comes, however, with the small coefficient
\begin{equation}
\frac{ \omega^2 \delta^2 }{ c^2 }
=
2 \rho \omega \varepsilon_0 
\sim 2 \times 10^{-12} \ll 1.
\label{eq:small-ratio-epsilons}
\end{equation}
This turns the prefactor into the convenient reference level
\begin{equation}
S^{\rm ref}_E = 
\frac{ k_{B} T \rho }{ 2\pi \, \delta^3 }
\sim 4 \times 10^{-17} \frac{ {\rm V^2/m^2} }{ {\rm Hz} }
\label{eq:spec-normalization}
\end{equation}
for good conductors (room temperature, 1\,MHz secular frequency).

\subsubsection{s-Polarization}
Negligible as mentioned above, discussed only for completeness.
The integral
\begin{equation}
s_s( d / \delta ) \approx
\int\limits_0^{\infty}\!{\rm d}x\,
\mathop{\rm Im}
\frac{ {\rm i} x - (2{\rm i} - x^2)^{1/2} }{
{\rm i} x + (2{\rm i} - x^2)^{1/2} }
 \, {\rm e}^{ - 2 x (d + t)/\delta }
\label{eq:s-integral-approx}
\end{equation}
goes for short distances, 
$d' = d + t \ll \delta$, to a constant because for $d' = 0$,
it converges to $2 / 3$.

\begin{figure}[htb]
\centerline{%
\includegraphics*[height=45mm]{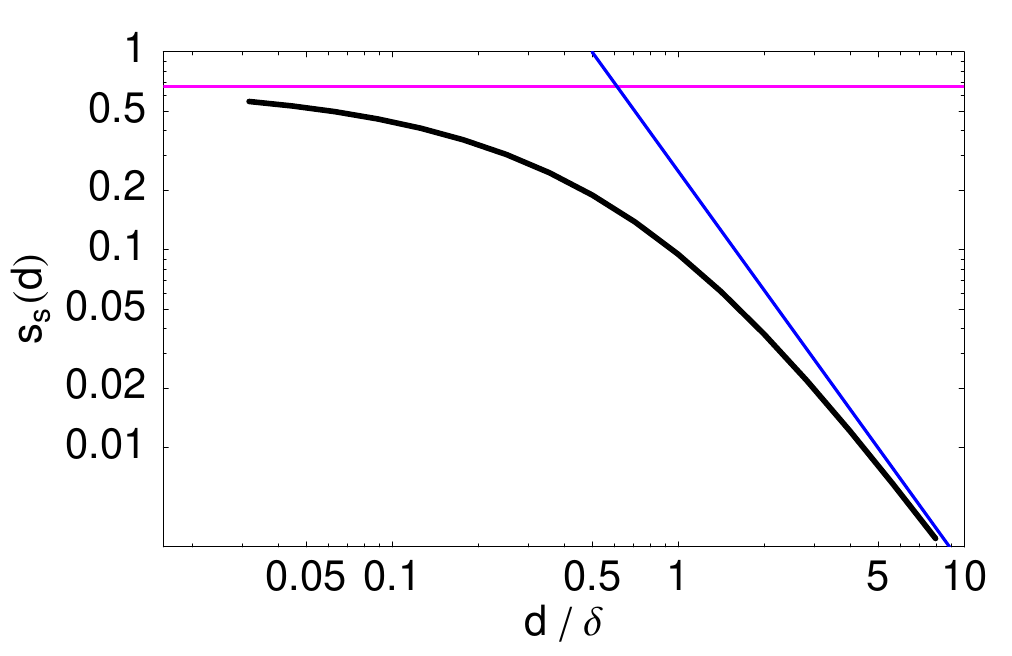}
\hspace*{2mm}
\includegraphics*[height=45mm]{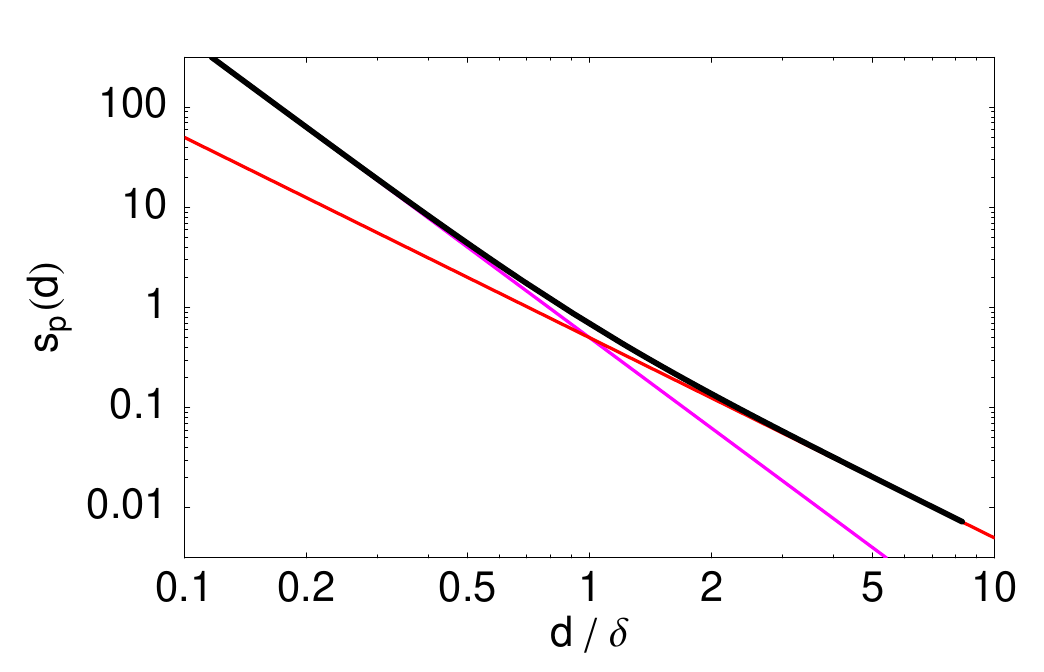}}
\caption[]{%
Dimensionless noise spectrum for electric fields, bare metal
surface. 
(\emph{left}) s-Polarization. Magenta: constant $2 / 3$;
blue: Eq.(\ref{eq:large-distance-s}) $\sim 1/d^2$.
(\emph{right}) p-Polarization (without the prefactor
$\varepsilon_0 \omega \rho$). Magenta: 
Eq.(\ref{eq:short-p}) $\sim 1/d^3$;
red: Eq.(\ref{eq:large-p}) $\sim 1/d^2$.
}
\label{fig:sp-polar}
\end{figure}

For large distances, $d' = d + t \gg \delta$, the domain $x \ll 1$
gives the dominant contribution. Expanding the integrand,
we get
\begin{equation}
s_s \approx
\int\limits_{0}^{\infty}\!{\rm d}x\,
x 
\,{\rm e}^{ - 2 x d'/\delta }
=
\frac{ 1 }{ 4\, (d' / \delta) ^2 }.
\label{eq:large-distance-s}
\end{equation}
The two asymptotes compare well with the numerical integration
(which is immediate to compute as well), as Fig.\ref{fig:sp-polar}
shows.

\subsubsection{p-Polarization}
Using the approximate form~(\ref{eq:rp-approx}) for the reflection coefficient, there are two contributions. One for the bare
metal surface (the expansion assumes that $d' = d + t \approx d$,
superscript m for `metal'):
\begin{equation}
s_p^{\rm m}( d / \delta; \omega ) =
2 \varepsilon_0 \omega \rho
\int\limits_0^\infty\!{\rm d}x\, x 
\mathop{\rm Im} ( 2 {\rm i} - x^2 )^{1/2}\,
{\rm e}^{ - 2 x d / \delta}.
\label{eq:p-integral-approx-m}
\end{equation}
For short distance, perform a large-$x$ expansion (exponential kept for
convergence) and obtain:
\begin{equation}
d \ll \delta: \qquad
s_p^{\rm m}( d / \delta; \omega ) \approx
2 \varepsilon_0 \omega \rho
\int\limits_0^\infty\!{\rm d}x\, x^2 \,
{\rm e}^{ - 2 x d / \delta}
=
\frac{ \varepsilon_0 \omega \rho }{ 2 (d/\delta)^3 }.
\label{eq:short-p}
\end{equation}
And for large distance, small $x$ one obtains:
\begin{equation}
d \gg \delta: \qquad
s_p^{\rm m}( d / \delta; \omega ) \approx
2 \varepsilon_0 \omega \rho 
\int\limits_0^\infty\!{\rm d}x\, x \,
{\rm e}^{ - 2 x d / \delta}
=
\frac{ \varepsilon_0 \omega \rho }{ 2 (d/\delta)^2 }.
\label{eq:large-p}
\end{equation}
See Fig.\ref{fig:sp-polar}(right) how accurate these asymptotes are.

Note that the p-polarization involves, at large distance, the same (small) prefactor as
the s-polarized contribution, since $(\omega\delta/c)^2 =
2 \varepsilon_0 \omega \rho$ [Eq.(\ref{eq:small-ratio-epsilons})]. 
In particular, at large distances, both polarizations are comparable so
that parallel and normal fields have the same noise spectrum (we use $d' \approx d$
for consistency with the expansion of $r_p^{012}$):
\begin{equation}
d \gg \delta: \qquad
S^{\rm m}_{E,p}( d, \omega ) =
S^{\rm m}_{E,n}( d, \omega )
\approx
\frac{ k_{\rm B} T \rho }{ 2\pi \varepsilon_0 \delta^3 }
\frac{ 1 }{ (d / \delta) ^2 }.
\label{eq:large-distance-bare}
\end{equation}

The other contribution depends on the properties of the 
dielectric layer. Here the integral is simpler (superscript
d for `dielectric layer'):
\begin{equation}
s_p^{\rm d}( d / \delta; \omega ) =
- 2 \mathop{\rm Im} \frac{ \varepsilon_0 }{ \varepsilon_1 }
\frac{ t }{ \delta } 
\int\limits_0^\infty\!{\rm d}x\, x^3 
{\rm e}^{ - 2 x d / \delta}
=
- \mathop{\rm Im} \frac{ \varepsilon_0 }{ \varepsilon_1 }
\frac{ t }{ \delta } 
\frac{ 3 }{ 4 (d/\delta)^4 }.
\label{eq:p-integral-approx-d}
\end{equation}
Even a layer of a few nanometers makes this the dominant 
contribution, as can be seen in the plots [Fig.\ref{fig:spec-2}].
We recover again the factor $1/2$ between parallel
and normal spectra
\begin{equation}
S_{E,p}( d, \omega ) = 
\frac12 S_{E,n}( d, \omega ) \approx
\frac{ k_{\rm B} T }{ 2 \pi \varepsilon_0 \omega \delta^3 }
\mathop{\rm Im} 
\left( \frac{ - \varepsilon_0 }{ \varepsilon_1 } \right)
\frac{ t }{ \delta } 
\frac{ 3 }{ 4 (d/\delta)^4 },
\label{eq:layer-spec}
\end{equation}
This expression is independent of the properties of the metallic substrate
because the skin depth $\delta$ drops out.

\subsection{Normalized Power Spectra for Common Materials}
\begin{figure}[tbh]
\centerline{%
\includegraphics*[width=75mm]{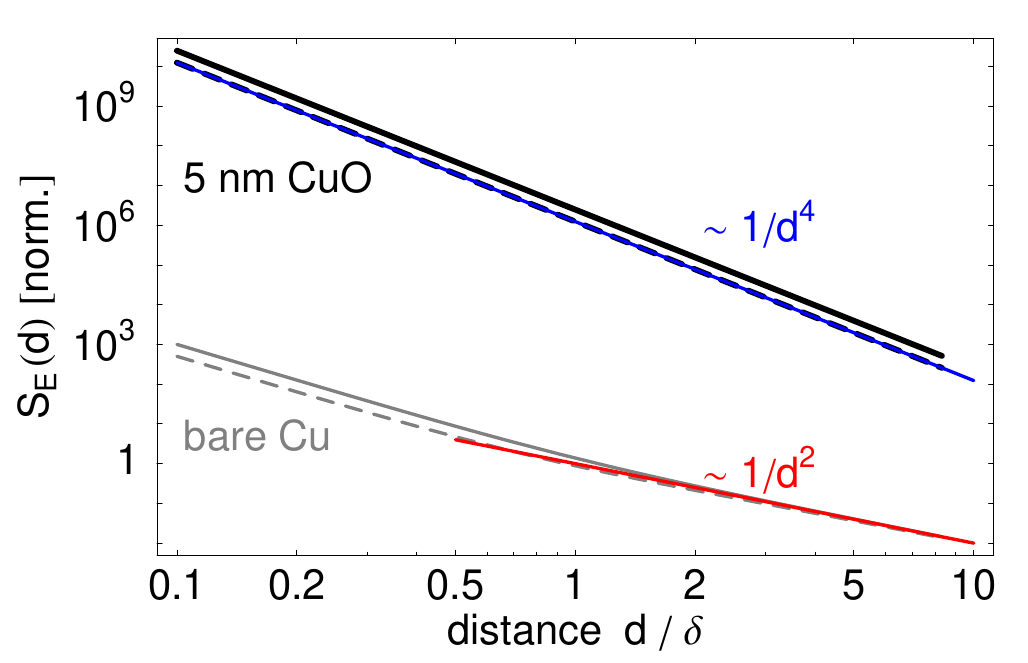}%
\hspace*{2mm}%
\includegraphics*[width=75mm]{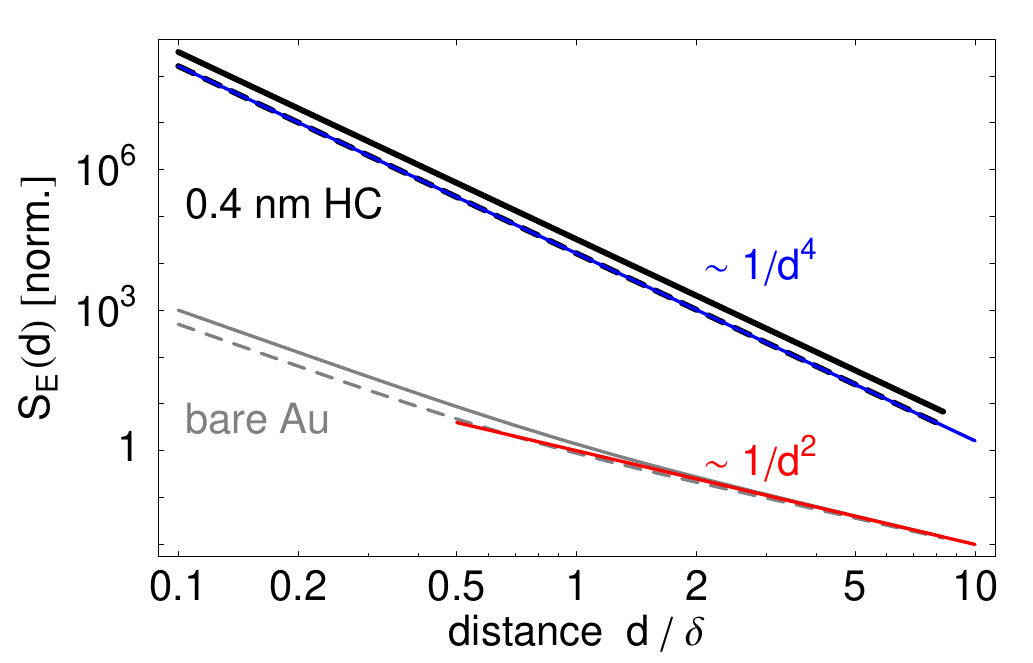}}

\vspace*{5mm}

\centerline{%
\includegraphics*[width=75mm]{spec-CuAu}}
\caption[]{%
Electric noise spectra vs.\ distance $d$ from 
bare and covered surfaces. Dashed: fields parallel to the surface;
solid: fields normal to surface. 
Thin red line: asymptote 
$\delta^2 / d^2$
;
thin blue line (coincident with parallel noise): 
asymptote $\sim 1/d^4$ of Eq.(\ref{eq:layer-spec}).
Top left: copper with conductivity from wikipedia
(skin depth $\delta = 65\,{\rm \mu m}$), 
oxide layer with $t = 5\,{\rm nm}$, dielectric constant $20$,
loss tangent $0.5$. Reference noise level 
[Eq.(\ref{eq:reference-noise-level})] is
$S_E^{\rm ref} \approx 4\times 10^{-17}\,{\rm (V/m)^2 / Hz}$. 
Top right: gold with wikipedia conductivity
(skin depth $\delta = 75\,{\rm \mu m}$), 
contamination layer (HC for hydrocarbon compounts) 
with $t = 0.4\,{\rm nm}$, 
dielectric constant $2$,
loss tangent $0.01$. Reference level 
$S_E^{\rm ref} \approx 3.5\times 10^{-17}\,{\rm (V/m)^2 / Hz}$. 
Bottom: comparison of both materials, identical to Fig.\ref{fig:noiseComparison}.}
\label{fig:spec-2}
\end{figure}

A convenient reference level of the noise is:
\begin{equation}
S_E^{\rm ref} 
= \frac{ k_{\rm B} T \times \rho \varepsilon_0 \omega 
	}{ 2\pi \varepsilon_0 \omega\,\delta^3 }
= \frac{ k_{\rm B} T \rho }{ 2\pi \,\delta^3 },
\label{eq:reference-noise-level}
\end{equation}
which is typical for a bare metal at distance $z \approx \delta$
(see caption Fig.\ref{fig:spec-2} for values).

The noise spectra parallel and perpendicular to the surface are 
shown in 
Fig.\ref{fig:spec-2}, 
normalized to Eq.(\ref{eq:reference-noise-level}). 
The thick lines give the results of 
numerical integrations, combining the contributions from
the metal and the dielectric layer and summing 
p- and s-polarization. 
The blue line, on top of the dashed black one, 
is given by the asymptote~(\ref{eq:layer-spec}) and depends
on the layer properties only.
One sees that the oxide coating (or the hydrocarbon contamination
layer) dominates the noise spectrum by more than 6 orders of
magnitude. For the bare metal, both p- and s-polarizations
contribute equally when the distance $z$ is larger than the skin depth
$\delta$ -- this doubling of the noise would not have been
found from strict electrostatics. Red line: large-distance
asymptote~(\ref{eq:large-distance-bare}) for the bare metal.

\section*{References}
\bibliography{main}

\end{document}